\documentclass[aps,prev,twocolumn,preprintnumbers,floatfix,nofootinbib]{revtex4}
\usepackage{graphicx}
\usepackage{mathrsfs}
\usepackage{amssymb}
\usepackage{verbatim}

\newcommand{\beq}{\begin{equation}}
\newcommand{\eeq}{\end{equation}}
\newcommand{\bea}{\begin{eqnarray}}
\newcommand{\eea}{\end{eqnarray}}

\newcommand{\gsim}{\lower.7ex\hbox{$\;\stackrel{\textstyle>}{\sim}\;$}}
\newcommand{\lsim}{\lower.7ex\hbox{$\;\stackrel{\textstyle<}{\sim}\;$}}

\newcommand{\drawsquare}[2]{\hbox{
\rule{#2pt}{#1pt}\hskip-#2pt
\rule{#1pt}{#2pt}\hskip-#1pt
\rule[#1pt]{#1pt}{#2pt}}\rule[#1pt]{#2pt}{#2pt}\hskip-#2pt
\rule{#2pt}{#1pt}}

\newcommand{\fund}{\raisebox{-.5pt}{\drawsquare{6.5}{0.4}}}

\newcommand{\antifund}{\overline{\fund}}

\def\stacksymbols #1#2#3#4{\def\theguybelow{#2}
    \def\vp{\lower#3pt}
    \def\sp{\baselineskip0pt\lineskip#4pt}
    \mathrel{\mathpalette\intermediary#1}}

\def\intermediary#1#2{\vp\vbox{\sp
     \everycr={}\tabskip0pt
     \halign{$\mathsurround0pt#1\hfil##\hfil$\crcr#2\crcr
              \theguybelow\crcr}}}

\def\comment#1{}

\def\u1x{U(1)_X}
\newcommand{\nc}{\newcommand}
\nc{\LL}{L}
\nc{\vv}{\tilde{v}}
\nc{\ccdot}{\!\cdot\!}
\nc{\gsm}{G_{SM}}
\nc{\vfive}{\mathbf{5}\oplus\mathbf{\overline{5}}}
\nc{\vten}{\mathbf{10}\oplus\mathbf{\overline{10}}}
\nc{\zhol}{Z^{\rm hol}}

\begin{document}

\preprint{MCTP-06-06,MIFP-06-11}

\title{Multi-Brane Recombination and Standard Model Flux Vacua}

\author{Jason Kumar$^{a}$}
\email{jkumar@physics.tamu.edu}
\author{James D. Wells$^{b}$}
\email{jwells@umich.edu}
\vspace{0.2cm}
\affiliation{
${}^a$ Physics Department, Texas A\&M University, College Station, TX 77843 \\
${}^b$ Michigan Center for Theoretical Physics (MCTP) \\
Department of Physics, University of Michigan, Ann Arbor, MI 48109}

\begin{abstract}
In a previous work, we noted that vacua with higher flux could be obtained
by recombination of a small set of hidden-sector branes, numbering up to the
K\"ahler degrees of freedom left to be fixed in the problem.
Here, we discuss the general construction of Type IIB Standard
Model flux vacua in which multiple branes participate in brane
recombination in the hidden sector.  We present several
models with 4, 5 and 10 branes in a stack. They are
complete with Pati-Salam gauge group, Standard Model
chiral matter, and a hidden sector that combines with nonzero
flux to satisfy RR tadpole constraints.  We also illustrate a
puzzle: within the 4-brane recombination approach,
we are unable to find a formal cutoff on
the number of Standard Model flux vacua.  Nevertheless, we show
that phenomenological
considerations are sure to cut off the number of vacua, albeit
at an extraordinarily large value. We comment on the possibility
that more formal constraints could further reduce the number
of vacua.

\end{abstract}

\maketitle

\maketitle

\setcounter{equation}{0}

{\it Benefits of Flux Vacua.}
Recently there has been much interest in flux vacua\cite{fluxvacua},
particularly in the question of how many flux
vacua one can construct which are ``phenomenologically
viable" -- vacua that are consistent with experimental knowledge.

Of course, it is extremely difficult to find any explicit
flux vacuum, let alone those that are phenomenologically
viable.  Instead, an alternative approach is to begin with
a class of flux vacua that retain certain phenomenologically
important properties, such as the gauge group and chiral matter
content of the supersymmetric Standard Model.
Each individual flux vacuum in this class will have moduli
with different vevs, resulting in different values for
the cosmological constant, as well as for other
phenomenological parameters such as ${\Lambda_{qcd}\over
M_{planck}}$, $M_{higgs}$, etc.  With a large enough number
of flux vacua, one can use statistics\cite{statistics} to estimate
the fraction of the flux vacua in this class
that have low-energy phenomenological parameters in a range
that is phenomenologically viable.  Results could lend support
to the existence of a stringy description
of quantum gravity with the correct details, even if an explicit
solution is lacking.

\bigskip

{\it Standard Model Embedding.} To this end, it is desirable to
find classes of flux vacua which are as large as possible, yet
contain Standard Model gauge group and matter content.  Some work in
this direction has already been done (\cite{IDB,MS}).  In
particular\cite{T6Z2Z2} there has been recent study of flux vacua of
Type IIB string theory compactified on an orientifold of $T^6/Z_2
\times Z_2$.  This model is noteworthy, because there is a simple
brane embedding using 3 D-brane stacks, such that gauge theory
living on these branes is a Pati-Salam left-right extension of the
Standard Model (i.e., $U(4) \times SU(2)_L \times SU(R)_R$), with
the same chiral matter content as the Standard Model\cite{MS}. The
branes composing these stacks are D9-branes with magnetic fields
turned on, inducing lower brane charge.
We will use magnetized
D-brane notation\cite{Cascales:2003zp},
where each brane is described by three co-prime ordered pairs
$(n,m)$.  Here, $m$ is the number of times the brane stack wraps
a torus, and $n={m\over 2\pi} \int F$ is the quantized magnetic
flux on the torus.
This embedding can be described as
\bea
N_a =8 &:& (1,0)(3,1)(3,-1) \nonumber\\
N_b=2 &:& (0,1)(1,0)(0,-1) \nonumber\\
N_c=2 &:& (0,1)(0,-1)(1,0),
\eea
where $N_i$ is the number of branes in the $i$th stack.
Notably, these branes are supersymmetric if the volumes
of the second and third tori are equal, ${\cal A}_2={\cal A}_3$.
The net D-brane charges of these stacks are
\beq
\{ Q_{D3},Q_{D7_1},Q_{D7_2},Q_{D7_3}\}=\{72,8,2,2\}
\eeq
where $D7_i$ refers to a D7-brane filling
space-time and wrapping each of the 3 compact
tori except the $i$th torus.

This brane embedding by itself
does not satisfy the RR-tadpole constraints (i.e., the
Gauss' law constraint that space-filling charge must
cancel):
\beq
\sum Q_{D3} =16,~~~
\sum Q_{D7_i} =16
\eeq
where the contribution on the right-hand-side is the
negative charge carried by the 64 O3- and 12 O7-planes.
As such, one must add ``hidden sector" charges
which can cancel these space-filling charges (and K-theory
tadpole constraints\cite{Ktheory}), as well as
provide phenomenologically desirable hidden sector
dynamics.  Fortunately, since RR/NSNS 3-form fluxes also
induce space-filling D3-brane charge
$L={1\over 2} \int F_{RR} \wedge G_{NSNS}$, these fluxes
can naturally be incorporated in this story.  These fluxes
contribute to the superpotential\cite{Gukov:1999ya},
which fixes the complex structure moduli (there are 51 of them
in this particular compactification) and the axio-dilaton.
Further non-perturbative corrections can fix the K\"ahler
moduli, leaving us with supersymmetric vacua with no
moduli and with Standard Model gauge group and matter content
in one sector.

\bigskip

{\it Importance of Large Flux.}
As shown in \cite{Denef:2004ze}, the  number of flux
vacua (i.e., the number of solutions to the $F$-term equations
generated by the GVW superpotential) will scale as
\beq
{\cal N}_{vac} \propto {L^{2n+2} \over n!},
\label{Nvac}
\eeq
where $n$ is the number of complex structure moduli.
This result is valid in the
limit $L\gg n$.  If  $L \sim n$, one instead finds the
scaling ${\cal N}_{vac}\sim
\exp {\sqrt{2\pi(2n+2)N_{flux}}}$~\cite{Ashok:2003gk}.

As such, we see that our task of finding classes of flux
vacua as large as possible is rephrased as the task
of finding compactifications with canceled RR-tadpoles
where the amount of flux we have turned on is as large as
possible.  Since we have fixed a choice of the visible sector, and
our orientifold contributes a fixed negative charge to the
tadpole constraints, this problem in turn is rephrased as that
of finding hidden sector branes which are supersymmetric, yet
contribute as negative a D3-brane charge as possible.  Indeed,
we see that (since $L$ is constrained
to be positive by the equations of motion) one cannot solve the
tadpole constraints at all unless the hidden sector branes contribute
negatively to the D3-brane tadpole condition.  But the more
negative this contribution, the more flux we can turn on to
cancel it.

In previous works, various examples of hidden sector branes have
been found which are supersymmetric and which contribute
negatively enough to the tadpole conditions to allow
flux to be turned on.  In the examples with the largest
amount of flux turned on\cite{Kumar:2005hf}, one relied on the process
of brane-recombination between two D-branes to generate a supersymmetric
hidden sector bound state with a more negative D3-brane charge than
otherwise would have been allowed.  In that earlier work, the
starting argument
was made that it would be possible to find flux vacua when one or two
additional NSNS tadpole constraints were added, since there were two degrees
of freedom among the K\"ahler moduli that could be tuned to accommodate
a supersymmetric solution.  Brane recombination relaxes that
conclusion.  Nevertheless, it was suggested that recombination involving
more than 2 D-branes would be unlikely to succeed in this toroidal model,
due to the difficulty in satisfying all RR-tadpole constraints.

The purpose of this article is to
point out that multi-brane recombination, in particular 4-brane combination,
can also be consistent with supersymmetric solutions.  Brane recombination
involving more than two D-branes can allow for even more negative
contributions to the hidden sector, thus permitting much
larger classes of flux vacua with Standard Model gauge group and
matter content. First, we review the basics of brane recombination
before illustrating the efficacy of multi-brane recombination in
producing phenomenologically viable solutions with high flux.

\bigskip

{\it Brane Recombination Basics.}
An undeformed D-brane preserves the same supersymmetry as the 
orientifold if it 
satisfies the NSNS tadpole constraint\cite{Berkooz:1996km}
\beq
\sum_{i=1}^{3} \tan^{-1}(m^a_i{\cal A}_i,n^a_i)=0~{\rm mod}~2\pi.
\eeq
But if this condition is not satisfied,
supersymmetry is not necessarily broken; instead, the branes
may deform, resulting in a supersymmetric bound state.  This
process is known as brane-recombination.  The result is that
we may end up with supersymmetric configurations even if we begin with
branes that do not satisfy the NSNS tadpole constraints.  This
gives us a much larger set of stable objects which we may use in
the construction of flux vacua.

In field theory language, the non-cancelation of the NSNS-tadpole
implies that there are non-zero
Fayet-Iliopoulos terms in the $N=1$ effective
supergravity\cite{Kachru:1999vj}.
For each $U(1)$ factor arising from the gauge theory on
a D-brane, there will be a D-term potential of the form:
\beq
V_{D_a} = {1\over 2g^2} \left(\sum q_i |\phi_i |^2 +\xi\right)^2
\eeq
where $\xi$ is the FI-term and the scalars $\phi_i$ arise
from strings stretching between branes and charged under this $U(1)$
(as well as another gauge group, generally).  We see that even
if the FI-terms are non-zero, scalars with appropriate charges
can become tachyonic and get vevs.  The total D-term potential
then vanishes, and supersymmetry is restored.  Of course, quantum
effects can correct the FI-term and the K\"ahler metric, but are
not expected to change the form of the D-term potential.

The question of whether or not this brane recombination process
can occur thus translates into the question of whether or not
there exist enough light scalars with appropriate charges for the D-term
potential to relax to zero.  Non-vectorlike scalars will arise from
strings stretching between two different branes (or between one brane
and the orientifold image of itself or another brane).  The number of
scalars will be determined by the topological intersection number
$I_{ab}=\prod_{i=1}^3 (n_i ^a m_i ^b -n_i ^b m_i ^a)$
between the branes $a$ and $b$.

The brane $a$ and its orientifold image $a'$
can intersect with any other brane $c$ or its orientifold
image $c'$, resulting in the following non-vectorlike matter
content:
\bea
ac~{\rm matter}: & ~ & I_{ac}~{\rm copies~of~}(\fund_a,\antifund_c)~{\rm chirals}
\nonumber\\
aa' ~{\rm matter}: & ~ & (I_{aa'}/2 -2I_{a,{\cal O}}  )~{\rm copies~ of}~sym.~{\rm chirals}
\nonumber\\
 & & (I_{aa'}/2 +2I_{a,{\cal O}}  )~{\rm copies~ of}~anti-sym.~{\rm chirals}
\nonumber\\
ac' ~{\rm matter}: & ~ & I_{ac'}~{\rm copies~ of}~ (\fund_a,\fund_c)~{\rm chirals }
\eea
where $I_{a,{\cal O}}$ is summed intersection number of $a$ with
each orientifold plane.

\bigskip

{\it A Four-Stack Model.}
We now present an explicit example of brane-recombination involving
4-stacks of D-branes:
\bea
N_q=2 &:& (-5,1)(-5,1)(-5,1) \nonumber\\
N_r=2 &:& (1,-1)(1,3)(1,3) \nonumber\\
N_s=2 &:& (1,3)(1,3)(1,-1) \nonumber\\
N_t=2 &:& (1,3)(1,-1)(1,3)
\eea
These branes each satisfy the NSNS tadpole constraints at
points in K\"ahler moduli space, but there exists no point where
all of them simultaneously satisfy the tadpole constraints.
The effective field theory will thus have non-zero FI-terms
($\xi \neq 0$).
The non-vectorlike spectrum arising from the open strings is
(in notation $(Q_q,Q_r,Q_s,Q_t)$) given by
\beq
\begin{array}{cc}
352(2,0,0,0)& 8(0,2,0,0) \\
8(0,0,2,0) & 8(0,0,0,2)  \\
1176(-1,-1,0,0) & ~~1176(-1,0,-1,0) \\
1176(-1,0,0,-1) & 24(0,-1,-1,0) \\
24(0,-1,0,-1) & 24(0,0,-1,-1) \\
1024(1,-1,0,0) & 1024(1,0,-1,0)\\
1024(1,0,0,-1) &
\end{array}
\eeq
There are
many scalars with either sign of charge under all
gauge groups; for any choice of signs of the FI-terms
there are enough scalars which become
tachyonic and can get vevs, causing the full D-term potential
to relax to zero.

One should worry that the superpotential Yukawa couplings
could generate a non-vanishing $F$-term.  However, we see that
brane recombination can occur in a very large subclass without
generating such $F$-terms.  For example, in the very large
volume limit we find $\xi_{q,r,s,t} \propto (-1,1,1,1)$.
Brane recombination can occur by only giving large vevs to scalars
with charges $(1,-1,0,0)$ and $(0,0,-1,-1)$, in which case gauge
invariance implies that no Yukawa coupling can yield dangerous
$F$-terms, and $V_F$ is parameterically small (note that this is not
a unique recombination method).

The total D-brane charges of
this bound state are:
\beq
\{Q_{D3},\vec Q_{D7_i}\}=\{ -244,\, 4,\, 4,\, 4\}
\eeq
This bound state can be part of the
hidden sector for the Pati-Salam left-right model
discussed earlier.  After brane-recombination, it will
preserve the same $N=1$ supersymmetry as the visible
sector and the orientifold plane.  The large negative
contribution means that we can also
turn on fluxes as part of the
hidden sector.  Since the flux contribution to
charge is quantized in units of 32 \cite{MS}, we
can add $N_{flux}=5$ units.  The corresponding number
of flux vacua for this model is\cite{Ashok:2003gk,Denef:2004ze}
\bea
{\cal N}_{vac} \sim 10^{25}  \times I
\eea
where $I$ is the integral of the vacuum density over
the complex structure moduli space (this is a fixed
number which depends only on the geometric data, not
the brane embedding).

\bigskip

{\it Varying Numbers of Stacks.}
One can easily see that this construction follows
through for varying numbers of stacks.  For
example, here are five stacks of D-branes
which may recombine to form a supersymmetric bound state:
\bea
N_q=4 &:& (-3,1)(-3,1)(-3,1) \nonumber\\
N_r=4 &:& (1,-1)(1,4)(1,4) \nonumber\\
N_s=4 &:& (1,4)(1,4)(1,-1) \nonumber\\
N_t=4 &:& (1,4)(1,-1)(1,4) \nonumber\\
N_u=2 &:& (-3,2)(-3,2)(-3,2)
\eea
The total D-brane charges of
this system are:
\beq
\{Q_{D3},\vec Q_{D7_i}\}=\{ -150,\, 4,\, 4,\, 4\}
\eeq
which would allow us to include this bound state as
part of the hidden sector, along with $N_{flux}=2$
unit of flux.

We do not go through the details of presenting the
spectrum and showing that brane-recombination can
occur, as this is merely a specific example of a
much more general class.  The point is that the
generic magnetized D-brane has non-trivial topological
intersection with another generic magnetized brane,
thus giving us a large number of charged scalars with
varying charges.  In general this is sufficient for
brane recombination to restore supersymmetry.

As another example, we consider a 10-stack brane
model:
\bea
N_q=4 &:& (2,-1)(2,5)(2,5) \nonumber\\
N_r=4 &:& (2,5)(2,-1)(2,5) \nonumber\\
N_s=4 &:& (2,5)(2,5)(2,-1) \nonumber\\
N_t=4 &:& (1,-1)(1,3)(1,3) \nonumber\\
N_u=4 &:& (1,3)(1,-1)(1,3) \nonumber\\
N_v=4 &:& (1,3)(1,3)(1,-1) \nonumber\\
N_w=2 &:& (-3,1)(-3,1)(-3,1) \nonumber\\
N_x=38 &:& (-2,1)(-2,1)(-2,1) \nonumber\\
N_y=50 &:& (-1,1)(-1,1)(-1,1) \nonumber\\
N_z=2 &:& (-4,1)(-4,1)(-4,1)
\eea
In this case, the total D-brane charges are:
\beq
\{Q_{D3},\vec Q_{D7_i}\}=\{ -428,\, 8,\, 8,\, 8\}
\eeq
and we can include the recombined bound state in
the hidden sector, along with $N_{flux}=11$
units of flux.  Indeed the counting arguments described
above suggest ${\cal N}_{vac} \sim 10^{37}\times I$ vacua,
which is
the largest number known in any explicit construction.

\bigskip

{\it An Abundance of Flux Vacua.}
This result shows that we can get large numbers of
flux vacua with a supersymmetric Standard Model
visible sector by utilizing bound states formed from
the recombination of multiple branes.  In fact, we
note a puzzle:
an {arbitrarily} large class of flux vacua with
Standard Model visible sector arising from constructions
in which the hidden sector has a bound state with
very large negative D3-brane charge, canceled by large
fluxes (related issues in IIA constructions are discussed
in \cite{IIA}; brane recombination allows one to evade
the bounds on flux found in \cite{Blumenhagen:2004xx}).

Consider the following generalization of our
4 brane stack:
\bea
N_q=2 &:& (-(x-1)^2,1)(-(x-1)^2),1)(-(x-1)^2,1) \nonumber\\
N_r=2 &:& (1,-1)(1,x)(1,x) \nonumber\\
N_s=2 &:& (1,x)(1,x)(1,-1) \nonumber\\
N_t=2 &:& (1,x)(1,-1)(1,x)
\eea
where $x$ is a large positive integer.  These branes will also
individually satisfy the NSNS tadpole constraints at points in
the moduli space, though not all at the same point (again yielding
$\xi \neq 0$).
We can similarly verify that there are enough scalars
with appropriate charges for tachyons to form and
condense, setting the D-term potential to zero.
The $N=1$ field theory analysis thus indicates that the bound
state formed after tachyon condensation is supersymmetric.
However, the net D-brane charge of this bound state will be
\beq
\{Q_{D3},\vec Q_{D7_i}\} =\{ -(x-1)^6 +3,\, 2,\, 2,\, 2\}
\eeq
Thus we see that as the integer $x$ is made arbitrarily
large, this bound state will contribute an arbitrary
negative charge to the D3-brane tadpole conditions, while
still not oversaturating the other D7-brane tadpole conditions.
This highly negative D3-brane contribution can be compensated by
a large amount of flux, yielding a large number of flux vacua.

This suggests the prospect of an arbitrary number of flux vacua
with a Standard Model visible sector.  If this were true,
then one might worry that there were an arbitrary number of
phenomenologically viable string vacua, which could be a
significant challenge to predictivity.

However, there is already reason to be suspicious of the field
theory analysis\footnote{We thank M. Douglas for discussions on this
point.}. The $D$-term potential basically computes the tree-level
worldsheet spectrum and verifies that there are enough tachyons to
lower all terms in the potential.  
This amounts to a stability analysis around the line of marginal 
stability parameterized in moduli space by $\xi_{FI} =0$.  One 
expects that this analysis will be exact in the limit where 
$\xi$ is small.  
However, in the cases we are interested in, the FI-terms can 
generically be string scale. This corresponds to having 
tachyons with string
scale masses, in which case it is not necessarily consistent to decouple
the excited string modes.  However, in similar contexts
the low energy effective field theory seems
to provide qualitatively correct answers even with string scale tachyons
\cite{lowEFT}.

The essential question is whether additional lines of marginal stability 
appear far in the interior of the moduli space.  If this occurs, one 
expects the above analysis to be correct (treating the FI-terms as adjustable 
parameters) until one crosses one of the new lines of marginal stability.

The bound state which we are considering has three positive charges
and one negative charge; as such, it satisfies the
criterion\cite{Blumenhagen:2004xx} for undeformed branes to be
supersymmetric somewhere in moduli space.  But the three positive
charges are all much smaller than the magnitude of the negative
charge.  If their charges were instead zero, then our state would
have the charge of an $\overline{D3}$-brane, and would thus break
supersymmetry.  One thus might suspect that there is indeed
a more subtle constraint on the ratio of charges which is being
missed by the D-term analysis.

Indeed, we can already see at least some additional lines of 
marginal stability must appear in the moduli space.  
A space-filling D3-brane preserves the
same supersymmetry as the orientifold planes.  In an $N=2$
compactification on the $T^6/Z_2 \times Z_2$ orbifold, the
central charge of this D3-brane is real and positive.
If a brane bound state preserves the same supersymmetry as
the orientifold, we expect that
its central charge must also be real and positive.  The D-term
analysis does not guarantee this; the D-term potential would not
notice if the bound state had negative central charge, provided the
contribution from fluxes and other branes is positive and large
enough to
cancel all other negative contributions.  
At large enough volume, however, the positive contribution to 
the central charge coming from the D7-branes will dominate.  Thus, 
an additional line of marginal stability must at least separate this 
large volume region from the region in which the central charge of 
the state become negative.

So although we suspect that many of the bound states found by 
the field theory analysis are stable, we know that many are not and 
we are not sure where the cutoff is.  But we can at least place a 
minimal cutoff on the number of bound states of this type by 
simply checking by hand if the brane bound state
in question has negative central charge.  The central charge is
given to lowest order by
\bea
Z = Q_{D3}+Q_{D7_1} {\cal A}_2 {\cal A}_3
+Q_{D7_2} {\cal A}_3 {\cal A}_1
+Q_{D7_3} {\cal A}_1 {\cal A}_2. \nonumber
\eea
where the ${\cal A}_i$ are real K\"ahler moduli (the volumes of the
various tori).

For the bound state in question, $Q_{D7_i}=2$.  If
$Q_{D3}$ becomes increasingly negative, $Z$ can only remain positive
if the volume of the tori become large.  But the sizes of these tori are
bounded by observation.  This places a
{\it phenomenological} bound on how negative $Q_{D3}$ can become.


The most generic bounds on the size of extra dimensions
arise from gravitational couplings, and
permit negative charges which are
extremely large (e.g., $|Q_{D3}|\lsim 10^{65}$).
The flux required to offset this large anti-D3 brane charge implies
from eq.~\ref{Nvac} an extraordinarily large number of flux vacua
(e.g., ${\cal N}_{vac}\sim 10^{6172}\times I$).

Note that we are not attempting to point out this particular model 
as an example of a string compactification with a very large 
number of vacua.  In fact, this particular string construction was 
chosen chiefly for its simplicity, and there are several defects 
relating to the construction of SM gauge theories which arise in 
this $T^6 /Z_2 \times Z_2$ construction.  In particular, in toroidal 
examples the constraints on gauge couplings will place very tight 
limits on the size of extra dimensions.  Furthermore, there are 
constraints on the cycles which NSNS fluxes may wrap in order to 
avoid anomalies on branes wrapping the same cycles\cite{Cascales:2003zp}.  
For toroidal 
models in which all branes wrap only a few cycles, this may create 
non-smooth distributions for the moduli which determine gauge couplings.
But these constraints are not likely to be limiting in intersecting 
brane models on more general orientifolds with more cycles in 
play, where bound states of the same nature as those described here 
can also arise.

One may also question the extent to which this entire discussion 
of brane stability is valid in a case where fluxes are turned on.  
In particular, one may wonder if the deformations of the geometry 
induced by backreaction to the fluxes negates the entire picture of 
branes wrapping cycles.  But there does not seem to be any reason 
to believe that this is the case.  The global topology of the 
orientifolded Calabi-Yau compactification is undisturbed by the 
addition of fluxes (once the RR-tadpoles are canceled).  If global 
tadpoles are canceled and local flux energy densities are low (which 
will be the case in the large volume limit of interest), there is no 
apparent reason for a brane to be unable to wrap a cycle; if field 
theory is appropriate, there is no reason why a supersymmetric field 
theory configuration (in the presence of flux energy densities) should 
fail to be stable.  Furthermore, it is amusing to note that we can 
always choose to replace our flux entirely by pure D3-branes.  In such 
a case, of course, we no longer worry about fixing complex structure 
moduli.  But we will see this very large number of field theoretic 
supersymmetric bound states (limited only by cutoffs on the size of 
the extra dimensions) with no complications arising from fluxes. 

As we allude to in the conclusions below,
the extraordinarily large numbers at play here strain our
sensibilities for the amount of charge a stable bound state
can carry.  Nevertheless, we have not yet encountered nor have
been able to devise
a rigorous argument that forbids these states.

\bigskip

{\it Conclusions.}
All of the flux vacua solutions we have discussed
have {\it Pati-Salam gauge group and SM chiral matter content}.
We have shown that brane recombination allows us to obtain
many more
non-zero flux solutions consistent with supersymmetry.
The various brane recombination pathways occurring in
the hidden sector have no direct negative impact on the
existence of a visible sector that contains the right gauge
groups and chiral matter content to be phenomenologically
viable.

We have also seen that multi-brane recombination can
lead to an extraordinarily large number of
solutions.
It is not clear that these solutions will survive a deeper
level of scrutiny, as such large negative $Q_{D3}$
brane charge and finite $Q_{D7_i}$ charges (and more
generally additional lines of marginal stability) could
lead to instabilities not captured by the basic analysis
we have performed here.

We emphasize that the considerations behind these constructions are
not expected to be unique to $T^6/Z_2 \times Z_2$.  Bound states of
this form may be expected to arise in generic constructions.
As such, it is quite important to understand what, if any, other formal
theoretical constraints may place further limits on how negative the
bound state charges can be.  One source of constraints
could be a generalized version of $\Pi$-stability\cite{PiStab}.  
This type of analysis is
expected to be valid throughout the moduli space, and thus even in
regimes where the FI-terms are large.
Unfortunately,
the technology required to apply this generalization to our framework
is not yet fully known.  In particular, it is not known how to formulate 
$\Pi$-stability in the $N=1$ framework of orientifolds, nor in the case 
where fluxes are turned on.
Nevertheless, it seems a promising avenue for the study of this interesting
puzzle.  
Putting aside the subtleties involving both orientifolds and fluxes, we 
have attempted to use the currently known $\Pi$-stability framework to 
determine if the kind of bound states which we have discussed here are 
unstable to decay into known decay products.  Using some specific examples, 
our preliminary analysis shows no instability in the bound states which 
appear stable from the field theory point of view.  Of course, this 
analysis is far from definitive, not only because of the aforementioned 
subtleties involving fluxes and orientifolds, but also because we can only 
check for stability against decay into a particular set of decay products, 
namely, the simple branes which form the bound state.  A complete stability 
analysis would require one to know all stable objects at a particular point 
in moduli space, and then use the formalism of $\Pi$-stability to determine 
all stable objects at every other point in moduli space, so that one could 
be sure that the putative bound state could not decay into any possible set 
of decay products.  It is not yet known how to conduct such an extensive 
stability analysis, even in $N=2$ cases.
 
The question ``how many string vacua can describe the real world to within 
experimental precision?" is very phenomenological in nature.  
It is fascinating to note that this question might only be accessed by the 
most formal aspects of string theory.

\bigskip

{\it Acknowledgments.}
We gratefully acknowledge A. Bergman, J. Distler,
S. Kachru, R. Reinbacher, E. Sharpe, G. Shiu, W. Taylor 
and U. Varadarajan for useful discussions.
We are especially
grateful to M. Douglas for our many discussions on these and related topics.
This work is supported in part by
NSF grant PHY-0314712, the Department of Energy, and the Michigan Center for
Theoretical Physics (MCTP).


\begin{thebibliography}{99}

\bibitem{fluxvacua}
  R.~Bousso and J.~Polchinski,
  JHEP {\bf 0006}, 006 (2000)
  [hep-th/0004134].
S.~B.~Giddings, S.~Kachru and J.~Polchinski,
Phys.\ Rev.\ D {\bf 66}, 106006 (2002)
[hep-th/0105097].
  A.~Maloney, E.~Silverstein and A.~Strominger,
  hep-th/0205316.
  B.~S.~Acharya,
  hep-th/0212294.
  S.~Kachru, R.~Kallosh, A.~Linde and S.~P.~Trivedi,
  Phys.\ Rev.\ D {\bf 68}, 046005 (2003)
  [hep-th/0301240].
There are numerous other works on this subject.  For
more detailed references, see
  M.~Grana,
  Phys.\ Rept.\  {\bf 423}, 91 (2006)
  [hep-th/0509003].
  J.~Kumar,
  hep-th/0601053.

\bibitem{statistics}
M.~R.~Douglas,
JHEP {\bf 0305}, 046 (2003)
[hep-th/0303194].
  J.~P.~Conlon and F.~Quevedo,
  JHEP {\bf 0410}, 039 (2004)
  [hep-th/0409215].
J.~Kumar and J.~D.~Wells,
Phys.\ Rev.\ D {\bf 71}, 026009 (2005)
[hep-th/0409218].
  K.~R.~Dienes,
  hep-th/0602286.




\bibitem{IDB}
R.~Blumenhagen, L.~Goerlich, B.~Kors and D.~Lust,
  JHEP {\bf 0010}, 006 (2000)
  [hep-th/0007024].
C.~Angelantonj, I.~Antoniadis, E.~Dudas and A.~Sagnotti,
  Phys.\ Lett.\ B {\bf 489}, 223 (2000)
  [hep-th/0007090].
G.~Aldazabal, S.~Franco, L.~E.~Ibanez, R.~Rabadan and A.~M.~Uranga,
  JHEP {\bf 0102}, 047 (2001)
  [hep-ph/0011132].
G.~Aldazabal, S.~Franco, L.~E.~Ibanez, R.~Rabadan and A.~M.~Uranga,
  J.\ Math.\ Phys.\  {\bf 42}, 3103 (2001)
  [hep-th/0011073].
D.~Cremades, L.~E.~Ibanez and F.~Marchesano,
  JHEP {\bf 0207}, 009 (2002)
  [hep-th/0201205].
R.~Blumenhagen, V.~Braun, B.~Kors and D.~Lust,
  hep-th/0210083.
A.~M.~Uranga,
  Class.\ Quant.\ Grav.\  {\bf 20}, S373 (2003)
  [hep-th/0301032].
D.~Lust,
  Class.\ Quant.\ Grav.\  {\bf 21}, S1399 (2004)
  [hep-th/0401156].
E.~Kiritsis,
  Fortsch.\ Phys.\  {\bf 52}, 200 (2004)
  [hep-th/0310001].
  T.~P.~T.~Dijkstra, L.~R.~Huiszoon and A.~N.~Schellekens,
  Phys.\ Lett.\ B {\bf 609}, 408 (2005)
  [hep-th/0403196].
  G.~Honecker and T.~Ott,
  Phys.\ Rev.\ D {\bf 70}, 126010 (2004)
  [Erratum-ibid.\ D {\bf 71}, 069902 (2005)]
  [hep-th/0404055].
C.~Kokorelis,
  hep-th/0410134.
  T.~P.~T.~Dijkstra, L.~R.~Huiszoon and A.~N.~Schellekens,
  Nucl.\ Phys.\ B {\bf 710}, 3 (2005)
  [hep-th/0411129].
M.~Cvetic, T.~Li and T.~Liu,
hep-th/0501041.
  F.~Gmeiner, R.~Blumenhagen, G.~Honecker, D.~Lust and T.~Weigand,
  hep-th/0510170.


\bibitem{MS}
F.~Marchesano and G.~Shiu,
Phys.\ Rev.\ D {\bf 71}, 011701 (2005)
[hep-th/0408059].
F.~Marchesano and G.~Shiu,
JHEP {\bf 0411}, 041 (2004)
[hep-th/0409132].

\bibitem{T6Z2Z2}
  R.~Gopakumar and S.~Mukhi,
  Nucl.\ Phys.\ B {\bf 479}, 260 (1996)
  [hep-th/9607057].
  S.~Forste, G.~Honecker and R.~Schreyer,
  Nucl.\ Phys.\ B {\bf 593}, 127 (2001)
  [hep-th/0008250].
M.~Cvetic, G.~Shiu and A.~M.~Uranga,
  Phys.\ Rev.\ Lett.\  {\bf 87}, 201801 (2001)
  [hep-th/0107143].
M.~Cvetic, G.~Shiu and A.~M.~Uranga,
  Nucl.\ Phys.\ B {\bf 615}, 3 (2001)
  [hep-th/0107166].
M.~Cvetic, I.~Papadimitriou and G.~Shiu,
  Nucl.\ Phys.\ B {\bf 659}, 193 (2003)
  [Erratum-ibid.\ B {\bf 696}, 298 (2004)]
  [hep-th/0212177].
M.~Cvetic and I.~Papadimitriou,
  Phys.\ Rev.\ D {\bf 67}, 126006 (2003)
  [hep-th/0303197].
M.~Cvetic, T.~Li and T.~Liu,
  Nucl.\ Phys.\ B {\bf 698}, 163 (2004)
  [hep-th/0403061].
M.~Cvetic, P.~Langacker, T.~j.~Li and T.~Liu,
  Nucl.\ Phys.\ B {\bf 709}, 241 (2005)
  [hep-th/0407178].
  C.~M.~Chen, T.~Li and D.~V.~Nanopoulos,
  Nucl.\ Phys.\ B {\bf 732}, 224 (2006)
  [hep-th/0509059].
  C.~M.~Chen, T.~Li and D.~V.~Nanopoulos,
  Nucl.\ Phys.\ B {\bf 740}, 79 (2006)
  [hep-th/0601064].
  C.~M.~Chen, T.~Li and D.~V.~Nanopoulos,
  hep-th/0604107.

\bibitem{Cascales:2003zp}
J.~F.~G.~Cascales and A.~M.~Uranga,
JHEP {\bf 0305}, 011 (2003)
[hep-th/0303024].

\bibitem{Ktheory}
A.~M.~Uranga,
Nucl.\ Phys.\ B {\bf 598}, 225 (2001)
[hep-th/0011048].
E.~Witten,
Phys.\ Lett.\ B {\bf 117}, 324 (1982).
E.~Witten,
JHEP {\bf 9812}, 019 (1998)
[hep-th/9810188].
  J.~Maiden, G.~Shiu and B.~J.~Stefanski,
  hep-th/0602038.

\bibitem{Gukov:1999ya}
  S.~Gukov, C.~Vafa and E.~Witten,
  Nucl.\ Phys.\ B {\bf 584}, 69 (2000)
  [Erratum-ibid.\ B {\bf 608}, 477 (2001)]
  [hep-th/9906070].

\bibitem{Denef:2004ze}
F.~Denef and M.~R.~Douglas,
JHEP {\bf 0405}, 072 (2004)
[hep-th/0404116].

\bibitem{Ashok:2003gk}
S.~Ashok and M.~R.~Douglas,
JHEP {\bf 0401}, 060 (2004)
[hep-th/0307049].

\bibitem{Kumar:2005hf}
  J.~Kumar and J.~D.~Wells,
  JHEP {\bf 0509}, 067 (2005)
  [hep-th/0506252].

\bibitem{Berkooz:1996km}
M.~Berkooz, M.~R.~Douglas and R.~G.~Leigh,
Nucl.\ Phys.\ B {\bf 480}, 265 (1996)
[hep-th/9606139].

\bibitem{Kachru:1999vj}
  S.~Kachru and J.~McGreevy,
  Phys.\ Rev.\ D {\bf 61}, 026001 (2000)
  [hep-th/9908135].

\bibitem{IIA}
  O.~DeWolfe, A.~Giryavets, S.~Kachru and W.~Taylor,
  hep-th/0505160.

\bibitem{lowEFT}
  P.~Binetruy, G.~Dvali, R.~Kallosh and A.~Van Proeyen,
  Class.\ Quant.\ Grav.\  {\bf 21}, 3137 (2004)
  [arXiv:hep-th/0402046].
  A.~Sen,
  Int.\ J.\ Mod.\ Phys.\ A {\bf 20}, 5513 (2005)
  [arXiv:hep-th/0410103].

\bibitem{Blumenhagen:2004xx}
R.~Blumenhagen, F.~Gmeiner, G.~Honecker, D.~Lust and T.~Weigand,
Nucl.\ Phys.\ B {\bf 713}, 83 (2005)
[hep-th/0411173].

\bibitem{PiStab}
  M.~R.~Douglas, B.~Fiol and C.~Romelsberger,
  JHEP {\bf 0509}, 006 (2005)
  [hep-th/0002037].
  M.~R.~Douglas, B.~Fiol and C.~Romelsberger,
  JHEP {\bf 0509}, 057 (2005)
  [hep-th/0003263].
  M.~R.~Douglas,
  J.\ Math.\ Phys.\  {\bf 42}, 2818 (2001)
  [hep-th/0011017].
For reviews, see
  M.~R.~Douglas,
{\it Prepared for ICTP Spring School on Superstrings and Related Matters, Trieste, Italy, 2-10 Apr 2001}.
  P.~S.~Aspinwall,
  hep-th/0403166.


















\end{thebibliography}
\end{document}